\documentclass[journal]{IEEEtran}

\ifCLASSINFOpdf
\else
\fi

\usepackage{float}
\usepackage{graphicx,amssymb,lineno}
\usepackage{subfigure}
\usepackage{amsmath,amsfonts,amssymb}

\usepackage[linesnumbered,ruled,vlined]{algorithm2e}
\usepackage{algorithmic}
\usepackage[usenames]{color}
\usepackage{multirow} 
\usepackage{amsmath}

\begin{document}

\title{QoS-aware Full-duplex Concurrent Scheduling \\for Millimeter Wave Wireless Backhaul Networks}

\author{Weiguang~Ding,
        Yong~Niu,~\IEEEmembership{Member,~IEEE,}
        Hao~Wu, ~\IEEEmembership{Member,~IEEE,}
        Yong~Li,~\IEEEmembership{Member,~IEEE,}
        and~Zhangdui Zhong,~\IEEEmembership{Senior Member,~IEEE}
\thanks{W.~Ding, Y.~Niu, H.~Wu, and Z. Zhong are with the State Key Laboratory of Rail Traffic
Control and Safety, the School of Electronic and Information Engineering, and Beijing Engineering Research Center of High-speed Railway Broadband Mobile Communications, Beijing Jiaotong University, Beijing 100044, China (e-mails:16120056@bjtu.edu.cn; niuy11@163.com; hwu@bjtu.edu.cn).}
\thanks{Yong~Li is with State Key Laboratory on
 Microwave and Digital Communications, Tsinghua National Laboratory for Information
 Science and Technology (TNLIST), Department of Electronic Engineering, Tsinghua
 University, Beijing 100084, China (e-mail: liyong07@tsinghua.edu.cn).}
\thanks{This study is supported by the Fundamental Research Funds for the Central Universities Grant 2016RC056, the State Key Laboratory of Rail Traffic Control and Safety (Contract No.~RCS2017ZT009 and No.~RCS2016ZT015), Beijing Jiaotong University, National Natural Science Foundation of China Grants 61725101, and Beijing Natural Science Foundation Grants L172020. It's also supported in part by the Chinese National Programs for High Technology Research and Development 863 project (No.2015AA016005) and the China Postdoctoral Science Foundation under Grant 2017M610040.}
}


\maketitle

\begin{abstract}
The development of self-interference (SI) cancelation technology makes full-duplex (FD) communication possible. Considering the quality of service (QoS) of flows in small cells densely deployed scenario with limited time slot (TS) resources, this paper introduces the FD communication into the concurrent scheduling problem of millimeter-wave (mmWave) wireless backhaul network. We propose a QoS-aware FD concurrent scheduling algorithm to maximize the number of flows with their QoS requirements satisfied. Based on the contention graph, the algorithm makes full use of the FD condition. Both residual self-interference (RSI) and multi-user interference (MUI) are considered. Besides, it also fully considers the QoS requirements of flows and ensures the flows can be transmitted at high rates. Extensive simulations at 60GHz demonstrate that with high SI cancelation level and appropriate contention threshold, the proposed FD algorithm can achieve superior performance in terms of the number of flows with their QoS requirements satisfied and the system throughput compared with other state-of-the-art schemes.
\end{abstract}


\IEEEpeerreviewmaketitle

\section{Introduction}
In the fifth generation (5G) mobile cellular network, due to the densification of small cells, the massive backhaul traffic becomes a significant problem \cite{Ultra-dense networks}, \cite{zhuyun}. Compared with the fiber based backhaul network, the wireless backhaul network in millimeter-wave (mmWave) bands also has huge bandwidth, and can provide a more cost-effective and flexible solution to this problem \cite{Wireless backhaul}. In the mmWave wireless backhaul network, directional antennas and beamforming techniques are often used to compensate for the high path loss \cite{Beamtraining2,Xiao1}. The directional communication can reduce the interference between different flows, and thus concurrent transmissions (i.e. spatial reuse) of flows become possible. Concurrent transmissions can significantly increase the system throughput \cite{mao}. However, the concurrent transmissions of multiple flows result in higher mutual interference, which will conversely degrade the system performance. Therefore, how to efficiently schedule the flows transmitted concurrently is worth to study and thus has attracted considerable interest from researchers \cite{zhuyun}, \cite{exclusive region}-\cite{Energy-Efficient}.

\begin{figure}[htbp]
  \begin{center}
  \includegraphics[width=5cm]{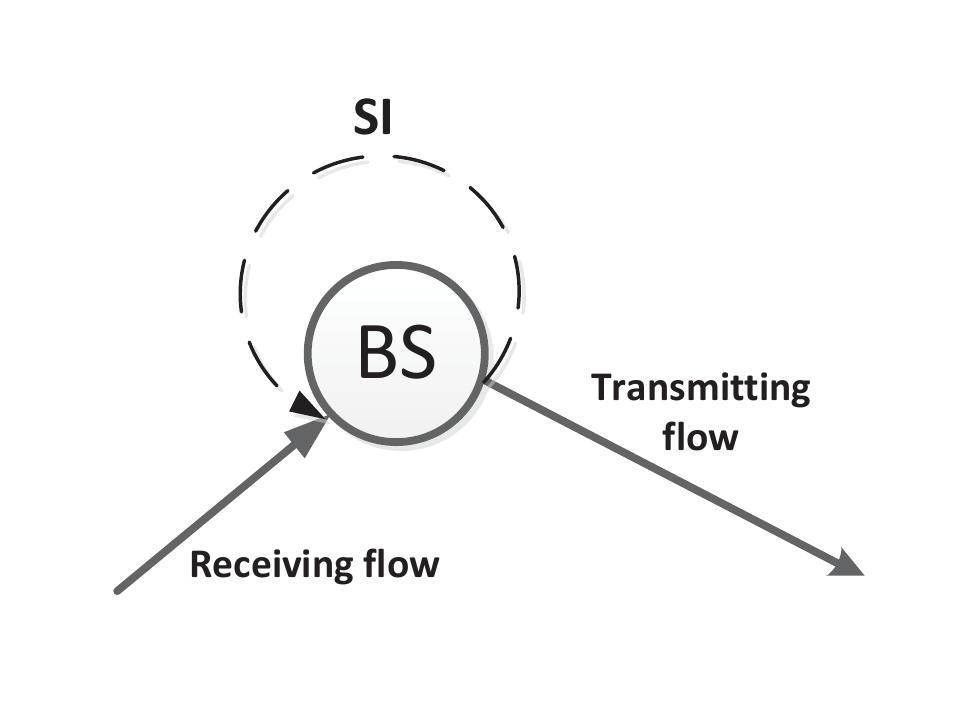}
  \caption{Self interference received at the same BS.} \label{fig:SI}
  \end{center}
\end{figure}

Most existing concurrent scheduling schemes \cite{zhuyun}, \cite{exclusive region}-\cite{Energy-Efficient} in mmWave bands hold the assumption of half-duplex (HD). Recently, with the development of self interference (SI) cancelation technology \cite{Practical}-\cite{Full-Duplex Millimeter-Wave Communication}, it becomes possible to enable the full-duplex (FD) communication in mmWave wireless backhaul networks \cite{Backhaul for 5G}. Here, the SI means the transmitted signal received by the local receiver at the same base station (BS) \cite{Full-Duplex Millimeter-Wave Communication}, which is shown in Figure \ref{fig:SI}. It seriously affects the performance of FD system \cite{Amplify-and-Forward Relaying}. By transmitting and receiving information simultaneously at the same BS over the same frequency \cite{Practical},\cite{Amplify-and-Forward Relaying}, the FD communication may theoretically double the spectral efficiency \cite{In-band full-duplex wireless}, which brings an important opportunity for the concurrent scheduling problem in mmWave wireless backhaul networks.

However, the SI can't be completely eliminated in practice. There is still residual self interference (RSI) in the system. Therefore, for the FD backhaul system, the interference we need to consider is more complex than that in HD system: not only multi-user interference (MUI), but also RSI. This is a big challenge for the concurrent scheduling problem in mmWave backhaul networks.

Moreover, in the future 5G mmWave backhaul network, many applications are bandwidth-intensive (e.g. uncompressed video streaming), and should be provided with multi-Gbps throughput \cite{STDMA}. The data flows of these applications all have their own minimum throughput requirements. In the following paper, the minimum throughput requirements will be referred to as the quality of service (QoS) requirements. To guarantee the required quality of service, the QoS requirements of flows need to be satisfied \cite{zhuyun}. Although in \cite{Backhaul for 5G}, Feng \emph{et al.} introduced the FD communication into the scheduling scheme for 5G mmWave backhaul networks, the scheduling solution was designed for the case with sufficient time slot (TS) resources. The QoS requirements were not specially considered in \cite{Backhaul for 5G}. Therefore, for the case where the TS resources are limited compared with the intensive traffic demands of users \cite{zhuyun},\cite{STDMA}, how to satisfy the QoS requirements of flows as many as possible is still a challenge.

The above opportunities and challenges motivate us to investigate a QoS-aware FD concurrent scheduling scheme for the mmWave wireless backhaul network with limited TS resources. The contributions of this paper can be summarized as follows.
\begin{itemize}
\item We innovatively introduce the FD technology into the concurrent scheduling problem of mmWave wireless backhaul networks with limited number of TSs. Both RSI and MUI are simultaneously taken into account so that the advantages of the FD technology and the concurrent transmission can be brought into full play.
\item The QoS requirements of flows in the case where the TS resources are limited are specially considered. We formulate a nonlinear integer programming (NLIP) problem aiming at maximizing the number of flows with their QoS requirements satisfied. Then, a QoS-aware FD scheduling algorithm is proposed, which can keep the flow rate high and satisfy the QoS requirements of flows as many as possible.
\item We evaluate the proposed algorithm in the 60GHz mmWave wireless backhaul network with limited TS resources. The extensive simulations demonstrate that compared with other state-of-the-art algorithms, the proposed QoS-aware FD algorithm can significantly improve the number of flows with their QoS requirements satisfied and the total system throughput. Furthermore, we also analyze the impact of SI cancelation level and contention threshold on the performance improvement.
\end{itemize}

The structure of this paper is organized as follows. Section II introduces the related work. Section III introduces the system overview and assumption. In Section IV, the optimal concurrent scheduling problem in FD mmWave wireless backhaul networks with limited TSs is formulated into an NLIP. In Section V, a QoS-aware FD concurrent scheduling algorithm is proposed. In Section VI, we conduct extensive simulations, and in Section VII we conclude this paper.
\section{Related Work}

Compared with the serial TDMA scheme, concurrent transmission scheduling can significantly increase the system throughput, and thus has been extensively studied \cite{zhuyun}, \cite{exclusive region}-\cite{Energy-Efficient}. Cai \emph{et al.} \cite{exclusive region} proposed a scheduling algorithm based on exclusive region to support concurrent transmissions. To maximize the number of flows scheduled in the network so that the QoS requirement of each flow is satisfied, Qiao \emph{et al}. \cite{STDMA} proposed a flip-based scheduling algorithm. In \cite{zhuyun}, Zhu \emph{et al.} proposed a Maximum QoS aware Independent Set (MQIS) based scheduling algorithm for mmWave backhaul networks to maximize the number of flows with their QoS requirements satisfied. In MQIS, the concurrent transmission and the QoS aware priority are exploited to achieve more successfully scheduled flows and higher network throughput. In \cite{stackelberg game}, based on Stackelberg game, Li \emph{et al} proposed a distributed transmission power control solution for the concurrent transmission scheduling between interference D2D links to further enhance the network throughput. Niu \emph{et al.} \cite{Energy-Efficient} proposed an energy efficient scheduling scheme for the mmWave backhaul network, which exploits concurrent transmissions to achieve higher energy efficiency. However, all the above scheduling algorithms assume the devices are HD.

Recently, the development of SI cancelation technology has made FD communication possible. Jain \cite{Practical} \emph{et al.} proposed the signal inversion and adaptive cancelation. Combining signal inversion cancelation with digital cancelation can reduce SI by up to 73dB. Everett \emph{et al.} \cite{Empowering} showed the BS could exploit directional diversity by using directional antennas to achieve additional passive suppression of the SI. Besides, Miura \emph{et al.} \cite{Node architecture} proposed a novel node architecture introducing directional antennas into FD wireless technology. Rajagopal \emph{et al.} \cite{SI mitigation} proved enabling backhaul transmission on one panel while simultaneously receiving backhaul on an adjacent panel is attainable for next generation backhaul designs. In \cite{Full-Duplex Millimeter-Wave Communication}, Xiao \emph{et al.} showed the configuration with separate Tx/Rx antenna arrays appeared more flexible in SI suppression, and proposed the beamforming cancelation in FD mmWave communication.

Considering the potential of the FD communication in increasing network performance, Feng \emph{et al.} \cite{Backhaul for 5G} proposed a design framework for 5G mmWave backhaul, which combined FD transmissions and hybrid beamforming with routing and scheduling schemes. However, the scheduling solution in \cite{Backhaul for 5G} was for the system with sufficient TS resources and aimed at accomplishing all of the transmissions with the minimum time. Thus, there was no special consideration for the QoS requirements of flows in limited time. Therefore, for mmWave backhaul networks with limited TS resources, a more QoS-favorable FD scheduling algorithm is needed.
\section{System Overview and Assumption} \label{section:III}

\begin{figure}[htbp]
  \begin{center}
  \includegraphics[width=9cm]{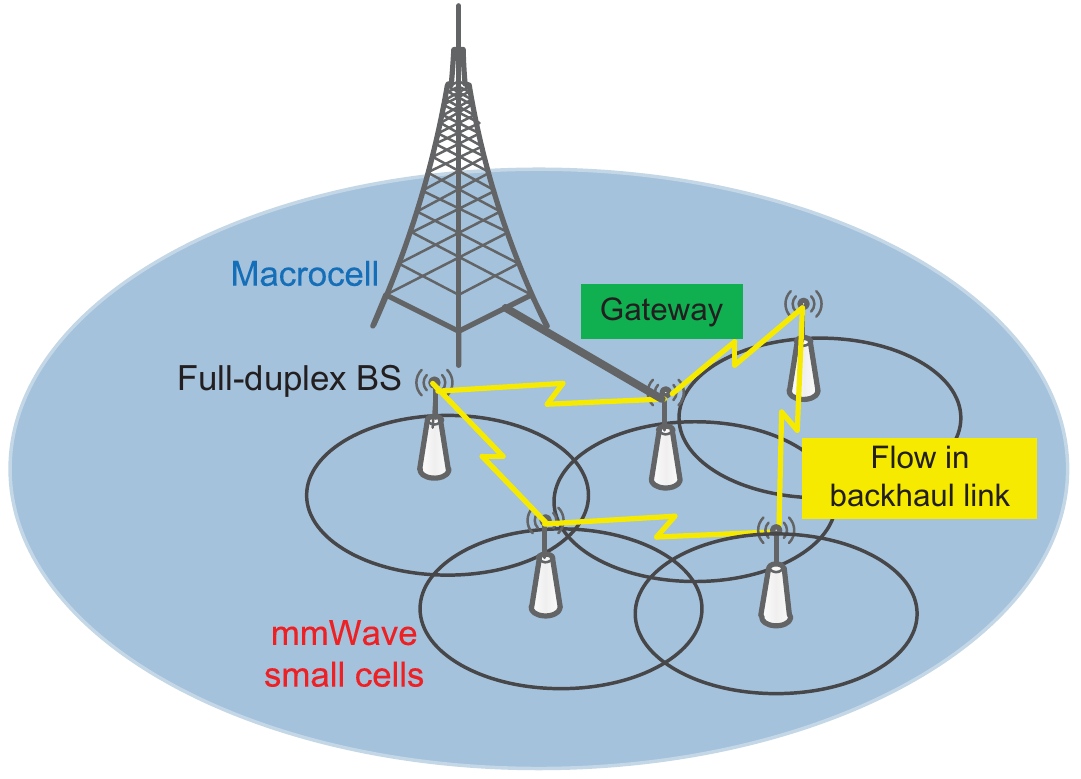}
  \caption{A full-duplex mmWave wireless backhaul network in the small cells densely deployed scenario.} \label{fig:smallcells}
  \end{center}
\end{figure}

\begin{figure}[htbp]
  \begin{center}
  \includegraphics[width=6cm]{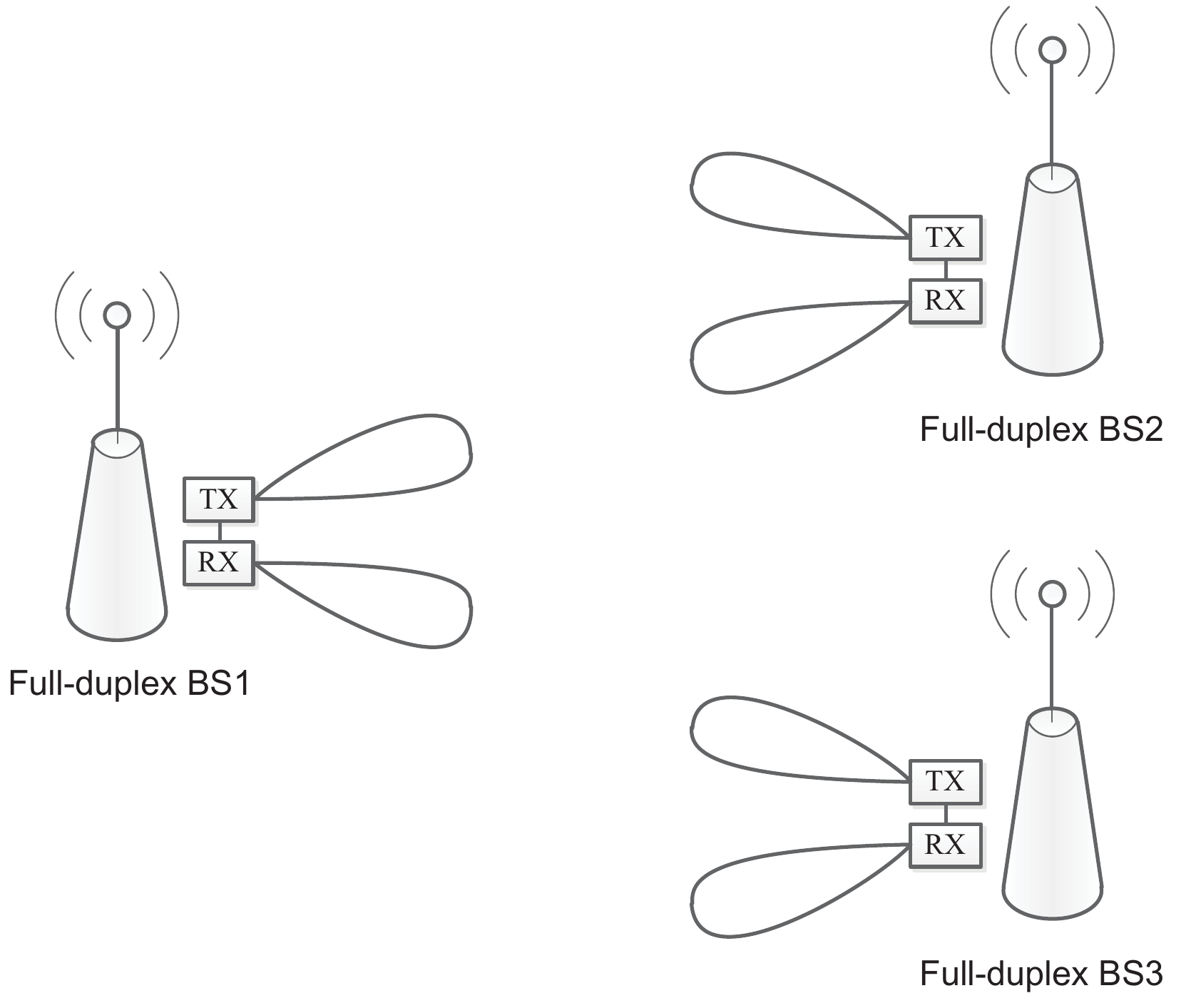}
  \caption{Full-duplex base stations.} \label{fig:FD_BS}
  \end{center}
\end{figure}

In this paper, we consider a typical FD mmWave wireless backhaul network in the small cells densely deployed scenario. As shown in Figure \ref{fig:smallcells}, the network includes $N$ BSs. The BSs are connected through backhaul links in the mmWave band. When there are some traffic demands from one BS to another, we say there is a flow between them. As shown in Figure \ref{fig:FD_BS}, each BS operates in FD mode and is equipped with two steerable directional antennas: one for transmitting and another for receiving. Therefore, a BS can at most simultaneously support two flows. It can simultaneously serve as the transmitter of one flow and the receiver of another, but it can't simultaneously serve as the transmitters or receivers of both two flows. There are one or more BSs connected to the backbone network via the macrocell, which is (are) called gateway(s) \cite{Energy-Efficient}. A backhaul network controller (BNC) resides on one of the gateways, which can synchronize the network, receive the QoS requirements of flows and obtain the locations of BSs \cite{survey}.
\subsection{The Received Power}
Since non-line-of-sight (NLOS) transmissions suffer from high attenuation, we use the line-of-sight (LOS) path loss model for mmWave as described in \cite{zhuyun}. For flow $f$, the received signal power at its receiver $r_f$ from its transmitter $t_f$ can be expressed as
\begin{equation}
P_r\left(t_f,r_f\right)=kP_tG_t(t_f,r_f)G_r(t_f,r_f)d_{t_fr_f}^{-n}.\label{equation: received signal power}
\end{equation}
$k$ is a factor that is proportional to $\left(\frac{\lambda}{4\pi}\right)^{\tiny{2}}$, where $\lambda$ denotes the wave length; $P_t$ denotes the transmission power of the transmitter; $G_t\left(t_f,r_f\right)$ denotes the transmitter antenna gain in the direction of from $t_f$ to $r_f$, and $G_r\left(t_f,r_f\right)$ denotes the receiver antenna gain in the direction of from $t_f$ to $r_f$; $d_{t_fr_f}$ denotes the distance between $t_f$ and $r_f$ and $n$ is the path loss exponent \cite{STDMA}.

According to the FD assumption mentioned above, the two flows scheduled simultaneously either have no common node or one's transmitter is the receiver of the other. Therefore, the interference between different flows can be divided into two cases: 1) the interference between two flows without any common node, namely, MUI; 2) the RSI after SI cancelation. The MUI caused by the transmitter $t_l$ of flow $l$ to the receiver $r_f$ of flow $f$ is defined as
\begin{equation}
P_r\left(t_l,r_f\right)=\rho kP_tG_t(t_l,r_f)G_r(t_l,r_f)d_{t_lr_f}^{-n},
\label{equation: received interference power}
\end{equation}
where $\rho$ is the MUI factor between different flows, which is related to the cross correlation of signals from different flows \cite{zhuyun}. 
According to \cite{Practical}, after SI cancelation, the effect of RSI can be modeled in terms of the SNR loss. Therefore, we can use $\beta_n N_0W$ to denote the RSI, where the non-negative parameter $\beta_n$ represents the SI cancelation level of the $n$th BS. The smaller $\beta_n$, the higher the level of SI cancelation. Due to various factors, we assume the parameters for different BSs are different. $N_0$ is the onesided power spectral density of white Gaussian noise; $W$ is the channel bandwidth.
\subsection{Data Rate}
With the reduction of multipath effect, mmWave channel can be approximated as Gaussian channel. With the interference from other flows, the data rate of flow \emph{f} can be estimated according to the Shannon's channel capacity \cite{Energy-Efficient}.

\subsection{Antenna Model}
In this paper, we adopt the realistic antenna model in \cite{Energy-Efficient}. The gain of a directional antenna in units of dB can be expressed as
\begin{equation}
G(\theta) =
\begin{cases}
G_0-3.01\times\left(\frac{2\theta}{\theta_{\mbox{\tiny{-3dB}}}}\right)^2, &\mbox{$0^{\circ}\le\theta\le\theta_{ml}/2$}\\
G_{sl}. &\mbox{$\theta_{ml}/2<\theta\le180^{\circ}$}
\end{cases}
\end{equation}
$\theta$ denotes an angle within the range $[0^\circ,180^\circ]$. The maximum antenna gain $G_0$ can be calculated as $G_0=\mbox{10log}(1.6162/\mbox{sin}(\theta_{{\mbox{\scriptsize{-3dB}}}}/2))^2$. $\theta_{{\mbox{\scriptsize{-3dB}}}}$ is the angle of the half-power beamwidth. The main lobe width $\theta_{ml}$ in units of degrees can be calculated as $\theta_{ml}=2.6\times\theta_{\mbox{\scriptsize{-3dB}}}$. The sidelobe gain $G_{sl}=-0.4111\times\mbox{ln}(\theta_{\mbox{\scriptsize{-3dB}}})-10.579$ \cite{Energy-Efficient}.
\section{Problem Formulation}
\begin{figure}[bp]
  \begin{center}
  \includegraphics[width=7cm]{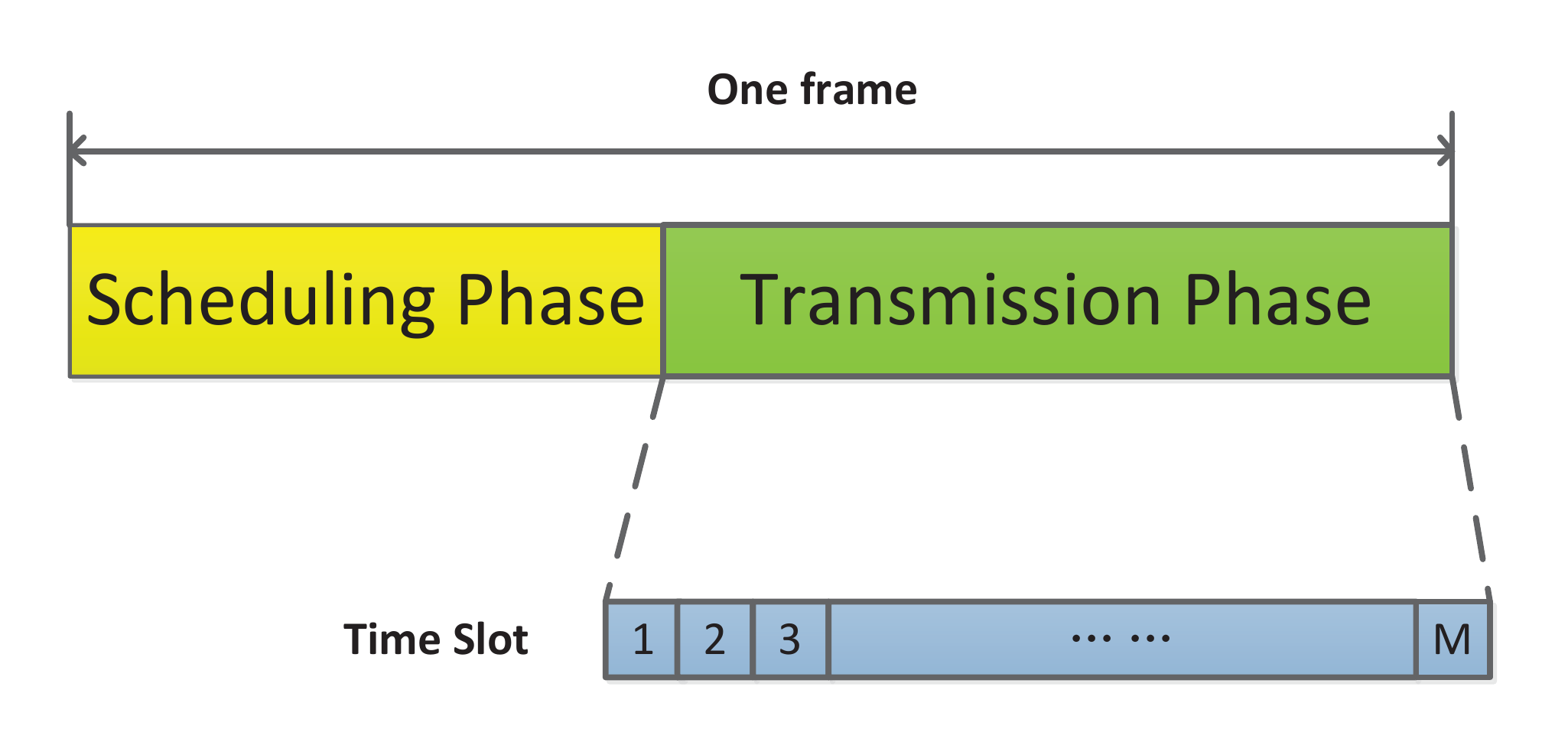}
  \caption{The structure of one frame.} \label{fig:frame}
  \end{center}
\end{figure}

In this paper, we consider a QoS-aware FD concurrent scheduling problem when the time is limited. System time is divided into a series of non-overlapping frames. As shown in Figure \ref{fig:frame}, each frame consists of a scheduling phase, where a transmission schedule \textbf{S} is computed by the BNC, and a transmission phase, where the BSs start concurrent transmissions following the schedule \cite{mao}. The transmission phase is further divided into $M$ equal TSs. It's assumed that there are $F$ flows in the network and each flow $f$ has its QoS requirement $q_f$. For each flow $f$, we define a binary variable $a_f^i$ to indicate whether flow $f$ is scheduled in the \emph{i}th TS. If so, $a_f^i = 1$; otherwise, $a_f^i = 0$. Since there may be different flows to be transmitted in different TSs, we denote the actual transmission rate of flow $f$ in the \emph{i}th TS by $R_f^i$. According to the Shannon's channel capacity \cite{Energy-Efficient}, $R_f^i$ can be calculated as

\begin{equation}
R_f^i=\eta Wlog_2(1+\frac{a_f^iP_r(t_f,r_f)}{N_0W+\sum\limits_{h}a_h^i\beta_{t_h} N_0W+\sum\limits_{l}a_l^iP_r(t_l,r_f)}).\label{equation:actual rate}
\end{equation}

where $\eta$ is the factor that describes the efficiency of the transceiver design, which is in the range of $(0,1)$. $W$ is the bandwidth, and $N_0$ is the one-sided power spectra density of white Gaussian noise. $h$ denotes the flow whose transmitter is the same as the receiver of flow $f$. $\beta_{t_h}$ is the parameter of SI cancelation level at BS $t_h$. $l$ denotes the flow without any common node with $f$. 

Then we can define the actual throughput of flow $f$ based on the schedule \textbf{S} as
\begin{equation}
T_{f}=\frac{\sum\limits_{i=1}^MR_f^i\triangle t}{T_s+M\triangle t},\label{equation:throughput}
\end{equation}
where $T_s$ is the time of scheduling phase and $\triangle t$ is the time of one TS. When the actual throughput $T_f$ of flow $f$ is greater than its QoS requirement $q_f$, we say flow $f$ has satisfied its QoS requirement, and call the flow a completed flow. A binary variable $I_f$ is used to indicate whether flow $f$ is completed. $I_f = 1$ indicates $f$ is completed, while $I_f = 0$ indicates $f$ is not completed.

As we investigate a QoS-aware scheduling for a backhaul network with limited time, given the QoS requirements of flows, with the limited number of TSs in the transmission phase, the optimal schedule should accommodate as many flows as possible \cite{zhuyun}. In other words, we should aim at maximizing the number of flows that satisfy their QoS requirements (i.e. the number of flows that are completed). Therefore, the objective function can be formulated as
\begin{equation}
\max\sum\limits_{f=1}^FI_f,
\end{equation}
and the first constraint is
\begin{equation}
I_f=
\begin{cases}
1, &\mbox{$T_{f} \ge q_f$;}\\
0, &\mbox{otherwise.}
\end{cases} \label{constraint 1}
\end{equation}

Next, we analyze the other constraints. Firstly, we use variable $f_n$ to denote the flow whose transmitter or receiver is the $n$th BS $B_n$, such as the transmitting flow and the receiving flow in Figure 1; thus $a_{f_n}^i$ indicates whether flow $f_n$ is scheduled in the $i$th TS, that is, whether $f_n$ does use $B_n$ in the $i$th TS. According to our FD assumption described in section \ref{section:III}, because each BS is just equipped with two steerable directional antennas, the number of flows that simultaneously use the same BS $B_n$ can't exceed two; this constraint can be expressed as

\begin{equation}
\sum\limits_{f_n}a_{f_n}^i\le2, \mbox{  }\forall i, n.
\end{equation}

Then we use $f_n^1$ and $f_n^2$ stand for the two flows that simultaneously use $B_n$; we also use $T (B_n)$ and $R (B_n)$ stand for the wireless links with $B_n$ as the transmitter and the receiver, respectively. As assumed in section \ref{section:III}, for the two antennas of a FD BS, one of them is a transmitting antenna and the other is a receiving antenna. Therefore, when two flows simultaneously use the same BS, the BS can only serve as the transmitter for one flow and as the receiver for the other, which can be expressed as:
\begin{equation}
\begin{split}
&f_n^1 \in T (B_n)\& f_n^2 \in R (B_n) \mbox{ }\\or&\mbox{ }f_n^1 \in R (B_n) \& f_n^2 \in T(B_n), \mbox{ if } \sum\limits_{f_n}a_{f_n}^i = 2.
\end{split} \label{constraint 3}
\end{equation}

In summary, the problem of optimal scheduling (\textbf{P1}) can be formulated as follows:
\begin{align*}
&\max\sum\limits_{f=1}^FI_f\\
&\mbox{s.t.}\\
&\mbox{Constraints (\ref{constraint 1}) - (\ref{constraint 3})}
\end{align*}

This is a nonlinear integer programming (NLIP) problem and is NP-hard \cite{zhuyun}. The optimization problem is similar to that in \cite{zhuyun}. \cite{zhuyun} is for the HD scenario while ours is for the FD scenario. Compared with [2], the number of constraints for our optimization problem is more, and the problem is obviously more complex than that in \cite{zhuyun}. \cite{zhuyun} is NP-hard, and thus our optimization problem is also NP-hard. In each TS, every flow is either scheduled or unscheduled. Therefore, when the number of TSs is $M$ and the number of flows is $F$, the computational complexity using exhaustive search algorithm is $2^{MF}$, which is exponential. In the small cells densely deployed scenario, the number of flows may be large, and thus it will be time-consuming if we use exhaustive algorithm to solve \textbf{P1}. The computational time is unacceptable for practical mmWave small cells where the duration of one TS is only a few microseconds \cite{Niu D2D}. Consequently, a heuristic algorithm with low complexity is desired to solve it in practice.
\section{QoS-aware Full-duplex Scheduling Algorithm}
In this section, we propose a QoS-aware full-duplex concurrent scheduling algorithm for problem \textbf{P1}. Borrowing the idea of contention graph from \cite{Energy-Efficient}, the algorithm makes full use of the FD condition and satisfies the QoS requirements of flows as many as possible. Next, we first describe how to construct the contention graph and then describe the proposed algorithm in detail.
\subsection{The Construction of Contention Graph}\label{Section: The Construction of Contention Graph}
In FD mmWave wireless backhaul networks, not all pairs of flows can be concurrently scheduled. In contention graph \cite{Energy-Efficient}, when the two flows can't be concurrently scheduled, we define there is a contention between them. In this paper, based on the assumption and analysis mentioned above, we define the flows that can't be concurrently scheduled into the following two cases.

\begin{figure}[htbp]
  \begin{center}
  \includegraphics[width=9cm]{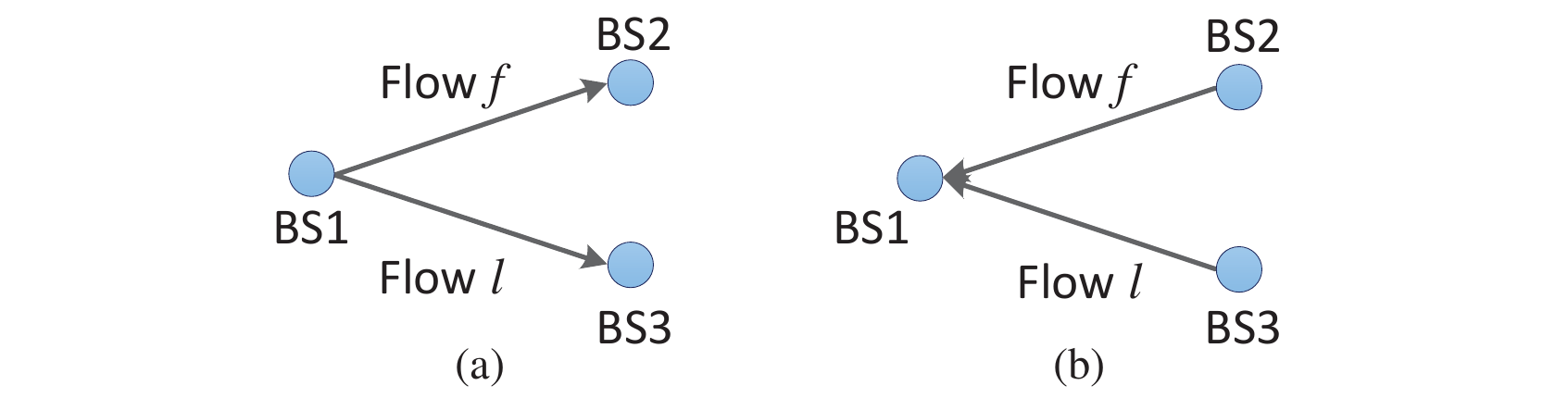}
  \caption{The flows that can't be concurrently scheduled due to the full-duplex assumption.}
  \label{fig:flows can't be simultaneously scheduled}
  \end{center}
\end{figure}

\begin{figure}[htbp]
  \begin{center}
  \includegraphics[width=9cm]{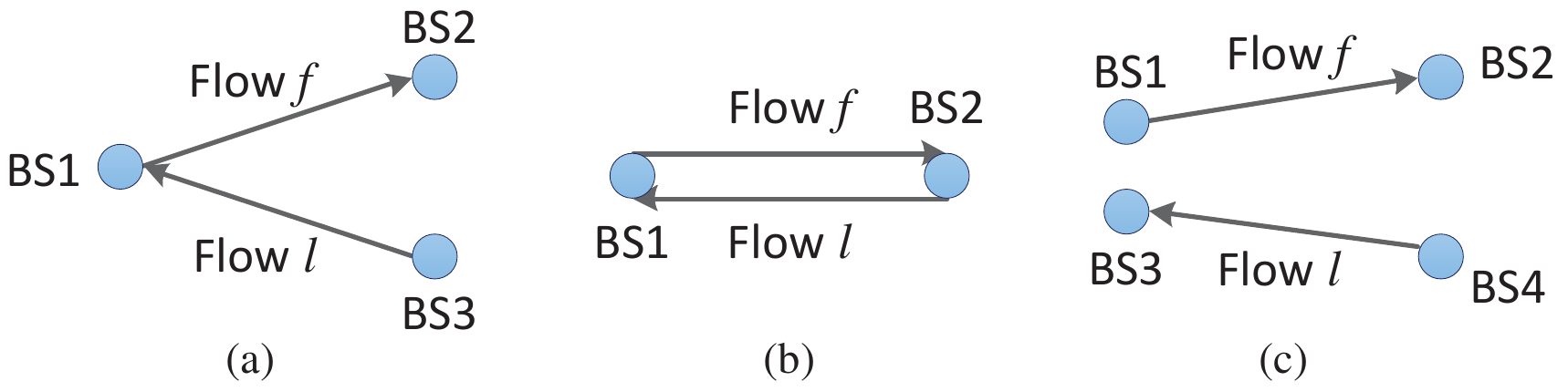}
  \caption{The flows that are allowed concurrently scheduled under the full-duplex assumption.}
  \label{fig:flows can be simultaneously scheduled}
  \end{center}
\end{figure}

Firstly, according to the FD assumption described in section \ref{section:III}, for the two antennas of a FD BS, one of them is a transmitting antenna and the other is a receiving antenna. Therefore, the two flows that simultaneously use the same BS as their transmitters (or receivers) can't be concurrently scheduled. This case is shown in Figure \ref{fig:flows can't be simultaneously scheduled}. Figure \ref{fig:flows can't be simultaneously scheduled} (a) shows that two flows simultaneously use the same BS as their transmitters. Similarly, Figure \ref{fig:flows can't be simultaneously scheduled} (b) shows that two flows simultaneously use the same BS as their receivers. Accordingly, based on the analysis for this case, the flows that can be concurrently scheduled are divided into following three cases. 1) As shown in Figure \ref{fig:flows can be simultaneously scheduled} (a), the transmitter of flow $f$ is the receiver of flow $l$, but the receiver of flow $f$ is not the transmitter of flow $l$. 2) As shown in Figure \ref{fig:flows can be simultaneously scheduled} (b), the transmitter of flow $f$ is the receiver of flow $l$, and the receiver of flow $f$ is the transmitter of flow $l$. 3) As shown in Figure \ref{fig:flows can be simultaneously scheduled} (c), the transmitter of flow $f$ is not the receiver of flow $l$, and the receiver of flow $f$ is not the transmitter of flow $l$, either.

Secondly, considering the QoS requirements of flows, to guarantee the flow rate and the system throughput, the two flows whose relative interference (RI) \cite{zhuyun} between each other is large can't be concurrently scheduled. When the RI between two flows is large, the rates of the flows become low. The low rates result in inefficient resource utilization. In other words, the TS resources are allocated to the flows, but the QoS requirements of them are hard to be satisfied, and thus they can't support the specific applications \cite{STDMA}. For the three cases in Figure \ref{fig:flows can be simultaneously scheduled}, we now define their RI, respectively. 1) For the case in Figure \ref{fig:flows can be simultaneously scheduled} (a), the interference from flow $f$ to flow $l$ is RSI. Therefore, the RI from flow $f$ to flow $l$ can be defined as
\begin{equation}
RI_{f,l} = \frac{N_0W+\beta_{t_f}N_0W}{P_r(t_l,r_l)}, \label{RI f to l under SI 1}
\end{equation}
where $P_r(t_l,r_l)$ is calculated as (\ref{equation: received signal power}). The interference from flow $l$ to flow $f$ is MUI; so the RI from flow $l$ to flow $f$ is defined as
\begin{equation}
RI_{l,f} = \frac{N_0W+P_r(t_l,r_f)}{P_r(t_f,r_f)}, \label{RI l to f under MUI}
\end{equation}
where $P_r(t_l,r_f)$ is calculated as (\ref{equation: received interference power}) and $P_r(t_f,r_f)$ is calculated as (\ref{equation: received signal power}).
2) For the case in Figure \ref{fig:flows can be simultaneously scheduled} (b), both the interference from flow $f$ to $l$ and that from flow $l$ to $f$ is RSI. Therefore, the RI between the two flows are both similar to (\ref{RI f to l under SI 1}).
3) For the case in Figure \ref{fig:flows can be simultaneously scheduled} (c), both the interference from flow $f$ to $l$ and that from flow $l$ to $f$ is MUI. Therefore, the RI between the two flows are both similar to (\ref{RI l to f under MUI}).

\begin{figure}[htbp]
  \begin{center}
  \includegraphics[width=4cm]{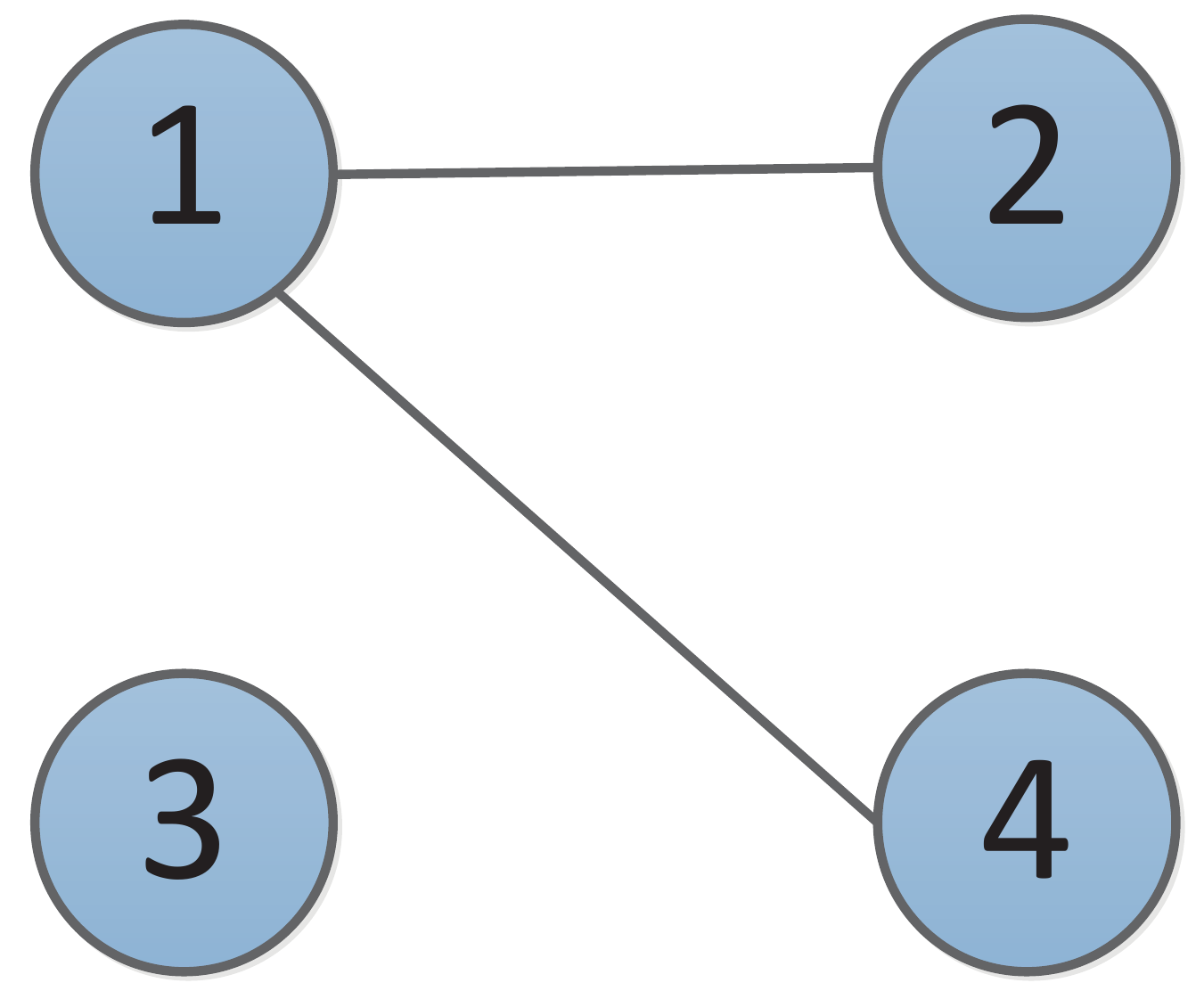}
  \caption{The contention graph with contention between flow 1 and flow 2 and
contention between flow 1 and flow 4.}
  \label{fig:contention graph}
  \end{center}
\end{figure}

Next, let's construct the contention graph. In the contention graph, each vertex represents a flow. If two flows can't be concurrently scheduled (i.e., there is a contention between them), an edge is inserted between the two corresponding vertices. For example, as shown in Figure \ref{fig:contention graph}, there is a contention between flow 1 and flow 2. In contrast, there is no contention between flow 1 and flow 3. Specifically, for the two pairs of flows in Figure \ref{fig:flows can't be simultaneously scheduled}, there is an edge between the two corresponding vertices, respectively. In addition, for the three pairs of flows in Figure \ref{fig:flows can be simultaneously scheduled}, we should examine whether the RI between the flows is too large. When the RI between two flows is larger than a contention threshold $\sigma$, we say there is a contention between them. In other words, if $\max (RI_{f,l},RI_{l,f})>\sigma$, an edge is inserted into the two corresponding vertices.

\subsection{The QoS-aware Full-duplex Scheduling Algorithm }
Based on the contention graph, we now concretely describe the proposed algorithm. The pseudo code for it is shown in Algorithm \ref{alg:The scheduling algorithm}.
\begin{algorithm}[htbp]
\DontPrintSemicolon
\caption{The Full-duplex Scheduling Algorithm} \label{alg:The scheduling algorithm}
BNC obtains Loc, $\beta_n$ and $q_f$;\\
calculate $\xi_f$ for each flow;\\
remove $\mathbb{D}=\{f|\xi_f>M\}$;\\
sort the remaining $F$ flows in non-decreasing order according to $\xi_f$ and get the pre-scheduling set $\mathbb{P}$;\\
generate \textbf{G} for all flows in $\mathbb{P}$;\\
\textbf{Initialization:} $\textbf{S}_{\scriptsize{F\times M}} =\textbf{0}$ and $change$ = 1; \\
\For {\rm{slot} $i\left(1\le i\le M\right)$}
{
\If {$change$ = \rm{1}}
{
\For {\rm{flow $f$} $\left(1\le f\le F\right)$}
{
\If {$\textbf{S}_i(f)=0$ \rm{and} $f$ \rm{has no contention with the flow(s) that is(are) ongoing}}
{
\If {\rm{set} $\textbf{S}_i(f)=1$ \rm{could increase the system throughput} }
{
$\textbf{S}_i(f)=1$ ;\\
}
}
}
}
$change$ = 0; $\textbf{S}_{i+1} = \textbf{S}_i $;\\
\If {\rm{any} $T_f > q_f$}
{
$change$ = 1;\\
$\textbf{S}_i(f)=-1$;\\
}
}
\end{algorithm}

To begin with, line 1 is some preparation work. The BNC obtains the BS location (Loc), the SI cancelation level ($\beta_n$) at each BS and the QoS requirement ($q_f$) of each flow. Next, in line 2, we calculate the number of TSs that each flow spends to complete its QoS requirement when there is no interference from others. The number of TSs is calculated as
\begin{equation}
\xi_f = \frac{q_f*\left(T_s+M\triangle t\right)}{R_f*\triangle t}.\label{equation: slot}
\end{equation}
$R_f$ is the rate of flow $f$ without interference from others, which can be calculated as
\begin{equation}
R_f=\eta Wlog_2(1+\frac{P_r(t_f,r_f)}{N_0W}).\label{equation:rate without interference}
\end{equation}

Since the scheduling problem we investigate is in limited time, i.e., in $M$ TSs, the flow whose $\xi_f$ is greater than $M$ should be removed. In the actual scheduling, there exists interference from other flows; so the flow rates will be further reduced and the spent number of TS will be further increased. The judgment for these flows becomes meaningless to our optimization goal. Removing the flows (represented by set $\mathbb{D}$) can not only reduce the complexity of subsequent scheduling, but also save more TSs to schedule more worthwhile flows, that is, the flows that can be completed in $M$ TSs. Pseudo code for this step is shown in line 3.

Then, as shown in lines 4-5, we sort the remaining $F$ flows in non-decreasing order according to $\xi_f$ and call the set of the sorted flows ``pre-scheduling set'' $\mathbb{P}$. Next, we construct the contention graph \textbf{G} for all the flows in  $\mathbb{P}$. A $F\times M$ matrix $\textbf{S}=[\textbf{S}_1,\textbf{S}_2,...\textbf{S}_i,... \textbf{S}_M]$ denotes the scheduling decision in $M$ slots. Vector $\textbf{S}_i$ indicates which flows are scheduled in the $i$th TS. If flow $f$ is scheduled in TS $i$, $\textbf{S}_i(f)$ = 1; if not, $\textbf{S}_i(f)$ = 0; if $f$ is completed, $\textbf{S}_i(f)$ = -1. Flag variable $change$ indicates whether there is(are) some flow(s) newly completed in every TS. If any, $change = 1$; if not, $change = 0$. For the first slot, $change$ is initialized to 1. The initialization steps are shown in line 6.

Then we make the scheduling decision slot by slot. In lines 9-12, to complete more flows in the limited TSs, we first determine the flow with the smallest $\xi$. In other words, we determine the flows in $\mathbb{P}$ one by one from the beginning. If flow $f$ has never been scheduled and has no contention with the flow(s) that is(are) ongoing, then the profit of scheduling the flow is evaluated: if scheduling it can increase the total system throughput, we schedule it; otherwise, skip the flow and determine the next. These rules help to guarantee the flow rate and the system throughput; so it's more QoS-aware.

In every TS, as shown in lines 14-16, it is necessary to check whether some flow(s) has(have) completed its(their) QoS requirement. If so, the corresponding $\textbf{S}_i(f)$ is(are) set to -1, which means the flow(s) will never be scheduled later. When one flow is completed, allocating resources to it is of little significance to further improve its QoS. Therefore, we should stop scheduling it and save the TSs to serve more other flows. At the same time, $change$ is set to 1.

In fact, as shown in line 8, only when it's the 1th TS or some flow(s) is(are) newly completed, that is, when $change$ = 1, we need to make new scheduling decision. If $change$ = 0, the scheduling vector is the same as the previous TS, which is shown in line 13. In this way, the scheduling complexity is greatly reduced. The algorithm is repeated until $M$ TSs are over, and we finally obtain the scheduling vector for each TS. Obviously, in the worst case, the variable \emph{change} is 1 in every TS, that is, there is(are) some flow(s) to be newly completed in every TS. Therefore, for the $M$ TSs and $F$ flows, the worst computational complexity of Algorithm \ref{alg:The scheduling algorithm} is $O(MF)$. 

\section{Performance Evaluation}
\subsection{Simulation Setup}
In the simulations, we evaluate the performance of the proposed algorithm in a 60GHz mmWave wireless backhaul network that 10 BSs are uniformly distributed in a $100m \times 100m$ square area. Every BS has the same transmission power $P_t$. The transmitters and receivers of flows are randomly selected, and the QoS requirements of flows are uniformly distributed between 1Gbps and 3Gbps. The SI cancelation parameters $\beta$ for different BSs are uniformly distributed in a certain range. To be more realistic, other parameters are shown in Table \ref{table:parameter setting}.
\begin{table}[bp]
\caption{Simulation parameters.} \label{table:parameter setting}
\centering  
\begin{tabular}{lccc}  
\hline
\textbf{Parameter} &\textbf{Symbol}&\textbf{Value}\\ \hline  
Transmission power &$P_t$ &1000mW\\
Path loss exponent  &$n$ &2\\
MUI factor &$\rho$ &1\\
Transceiver efficiency factor &$\eta$ &0.5\\
System bandwidth &$W$ &1200MHz\\
Background noise &$N_0$ &-134dbm$/$MHz\\     
Slot time &$\triangle t$ &18us\\
Scheduling phase time &$T_s$ &850us\\
Number of slots in transmission phase &$M$ &2000\\
Half-power beamwidth &$\theta_{\mbox\scriptsize{-3dB}}$ &$30^\circ$\\
 \hline
\end{tabular}
\end{table}

Because we focus on the QoS of flows, according to our optimization goal, we use the number of completed flows and the system throughput as evaluation metrics. When one flow achieves its QoS requirement, it is called a completed flow. System throughput represents the throughput of all flows in the network per slot.

To show the advantages of the proposed QoS-aware FD concurrent scheduling algorithm (Proposed-FD) in the network system with limited TS resources, we compare it with the following four schemes.

1) \textbf{TDMA}: In TDMA, the flows are transmitted serially. We use TDMA as the baseline for evaluating performance without concurrent transmissions.

2) \textbf{MQIS}: MQIS \cite{zhuyun} is a HD concurrent scheduling algorithm based on the maximum QoS-aware independent set. It first schedules the flow with the smallest degree in contention graph. It doesn't remove the flow(s) spent too much slots, nor does it evaluate the profits when adding a new flow. To the best of our knowledge, MQIS achieves the best performance in the network system with limited TS resources in terms of the number of completed flows and system throughput among the existing scheduling algorithms. Therefore, we use it as the baseline for evaluating performance without FD communication.

3) \textbf{Proposed-HD}: It uses the same scheduling algorithm with Proposed-FD, but it only allows the HD communication. We also use it as a baseline without the FD communication.

4) \textbf{FDP}: Full-Duplex (FDP) scheme \cite{Backhaul for 5G} is for the system where the TS resources are sufficient. It aims at accomplishing all of the transmissions with the minimum time. In every phase, higher priority is given to the flow that occupies the most TSs. If another flow is qualified to be transmitted together in the current phase, i.e., the number of flows simultaneously using the same BS doesn't exceed the number of RF chains and the SINR is larger than a certain threshold, the corresponding flow is also scheduled. Only when all the flows scheduled together in one phase are completed, can we start the next phase and make a new scheduling decision. We use it as a baseline for FD communication.

Each simulation performs 100 times to get a more reliable average result.
\subsection{Simulation Results}
\subsubsection{Under different numbers of flows}
\begin{figure}[t] 
\begin{minipage}[t]{1\linewidth}
\centering
\includegraphics[width=0.9\linewidth]{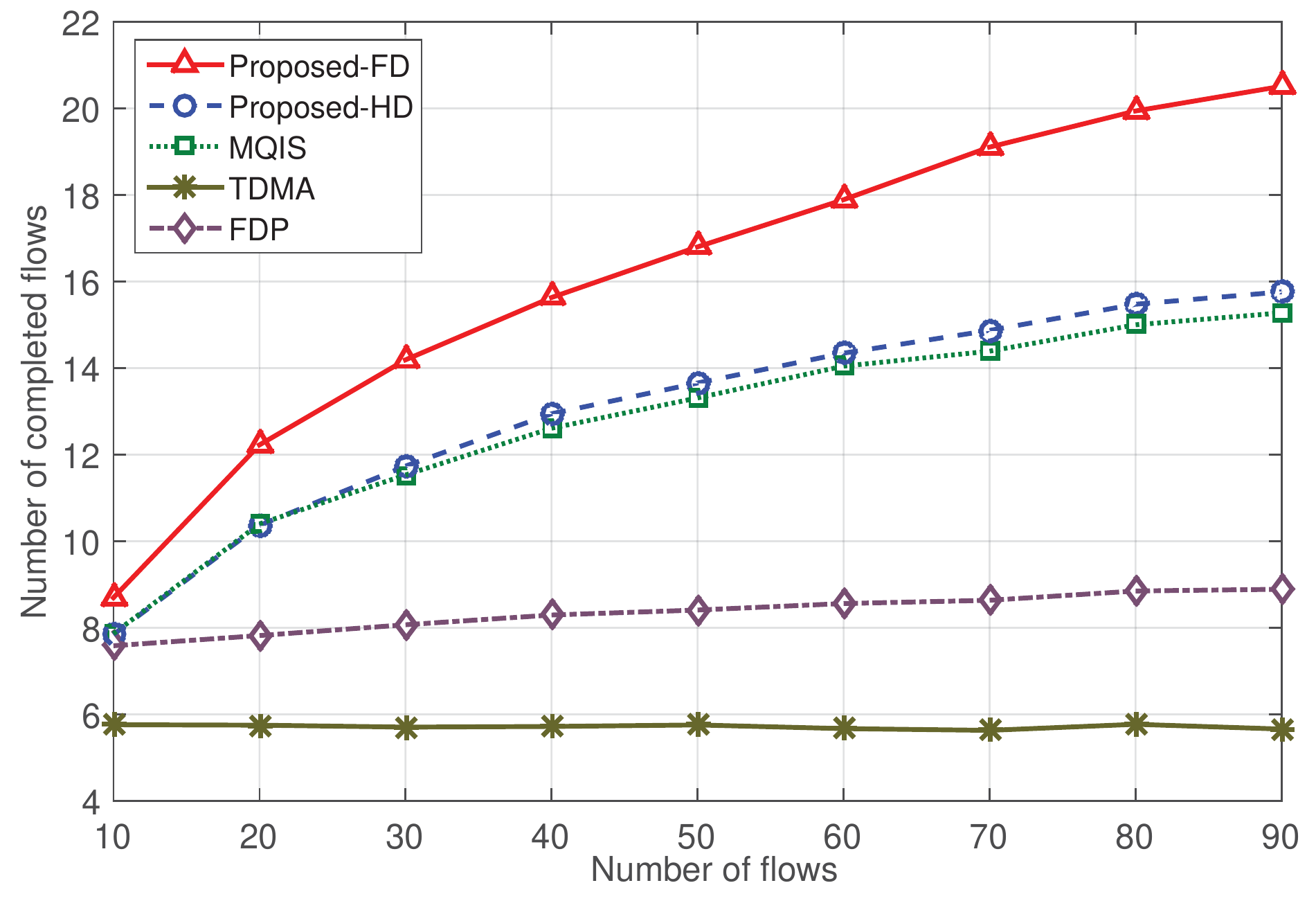}
\centerline{\small (a) number of completed flows}
\end{minipage}\\
\begin{minipage}[t]{1\linewidth}
\centering
\includegraphics[width=0.9\linewidth]{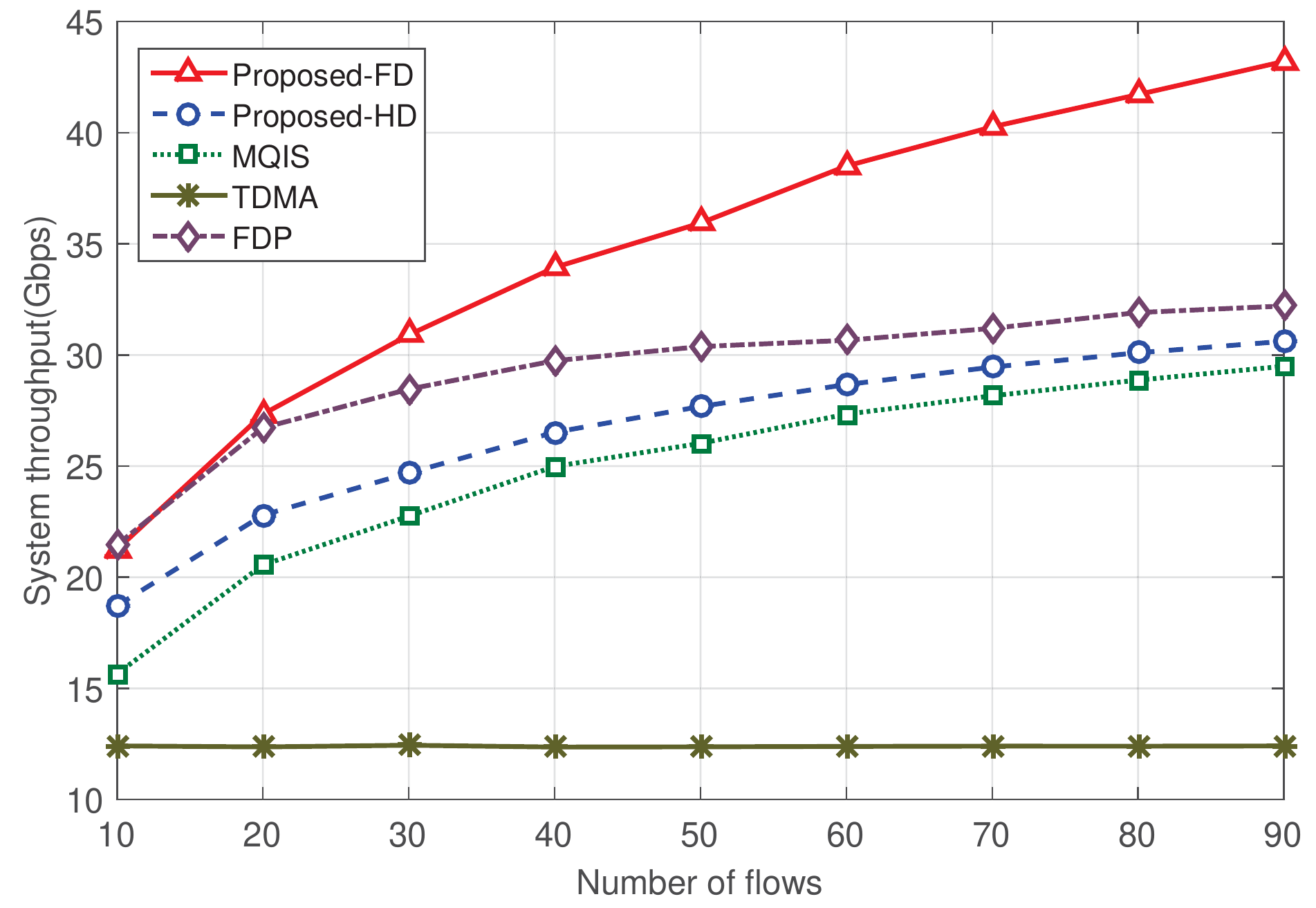}
\centerline{\small (b) system throughput}
\end{minipage}
\caption{The number of completed flows and system throughput under different number of flows.} \label{fig:under different number of flows}
\vspace*{-3mm}
\end{figure}

\begin{figure}[tbp] 
\begin{minipage}[t]{1\linewidth}
\centering
\includegraphics[width=0.9\linewidth]{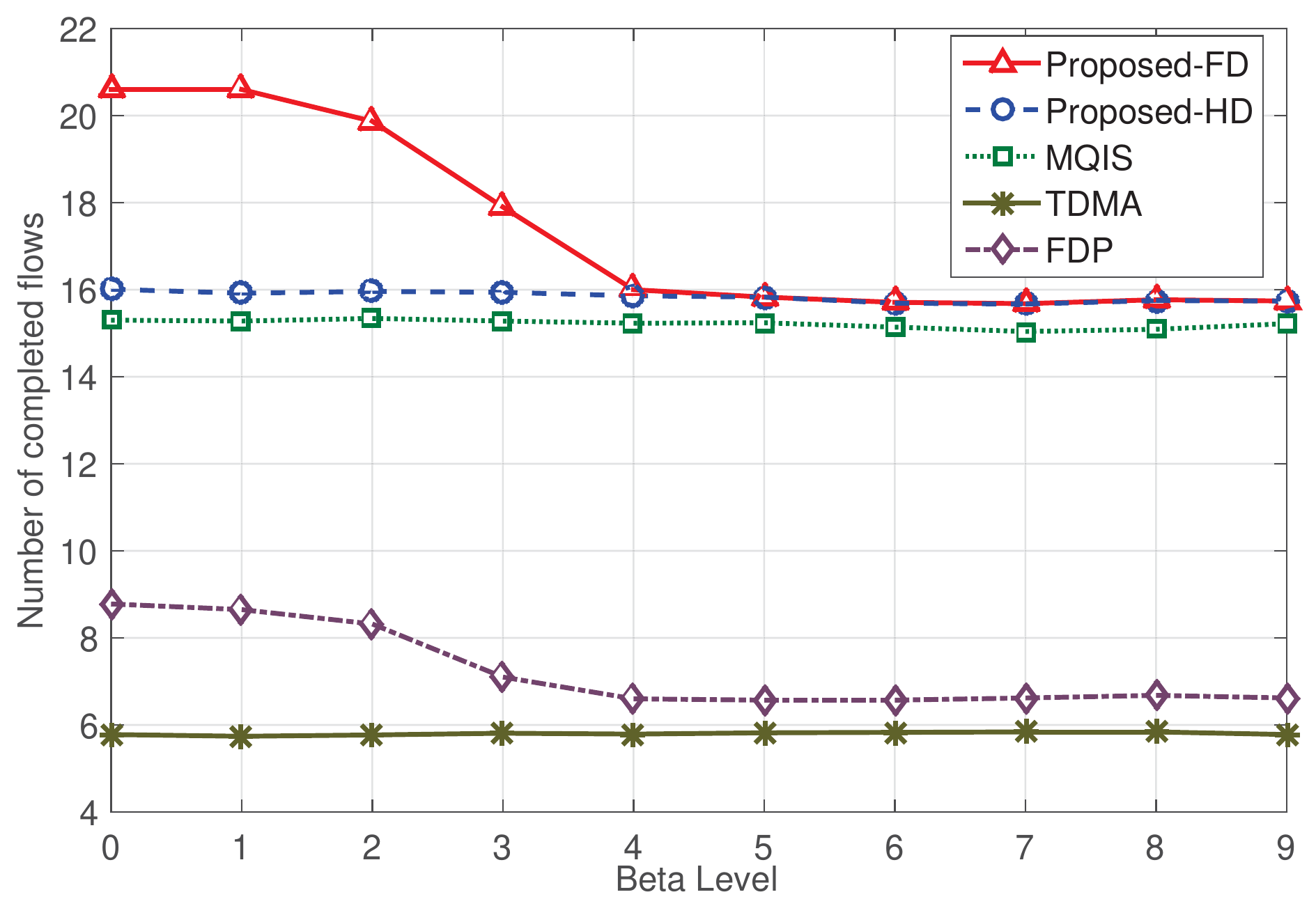}
\centerline{\small (a) number of completed flows}
\end{minipage}\\
\begin{minipage}[t]{1\linewidth}
\centering
\includegraphics[width=0.9\linewidth]{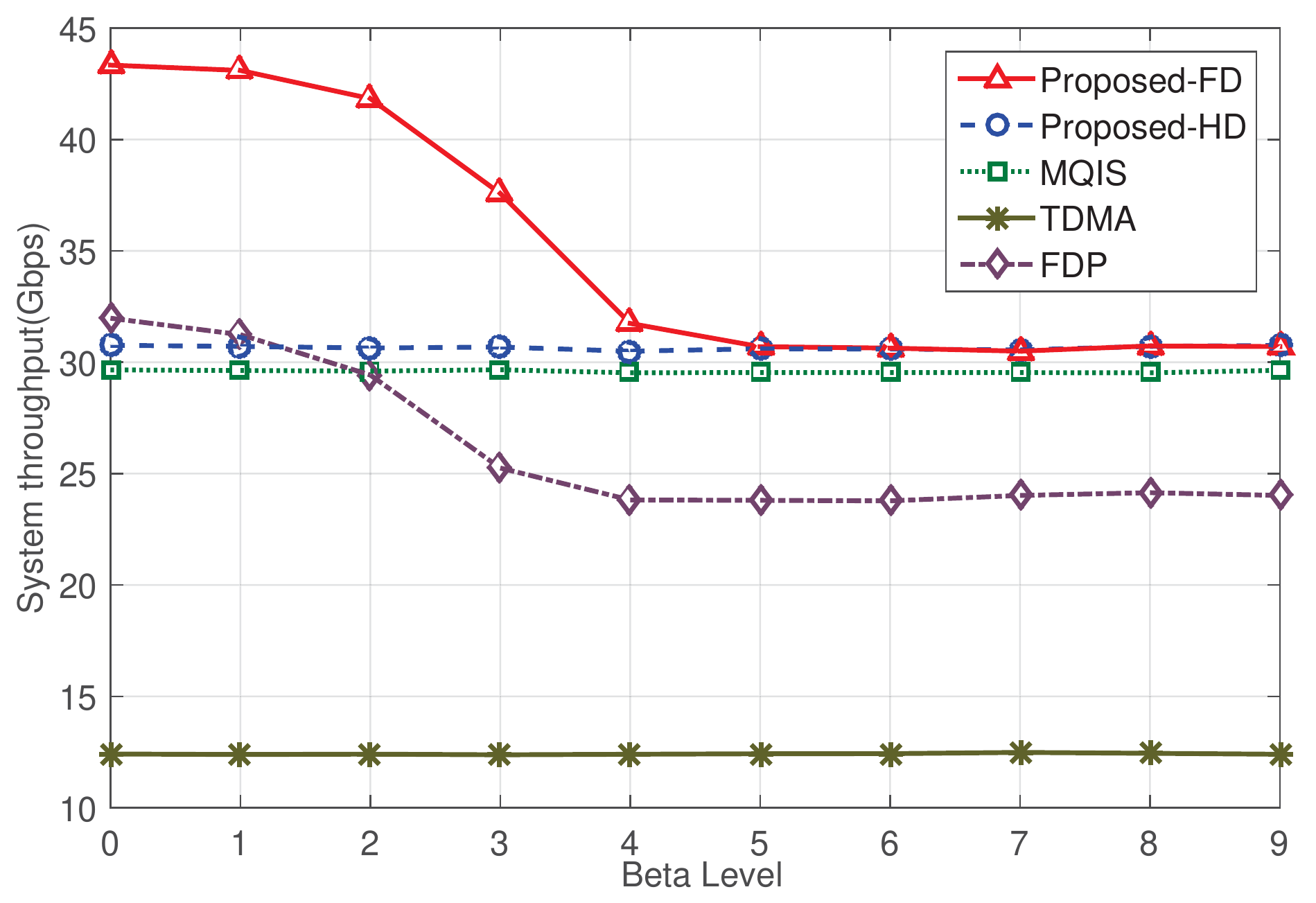}
\centerline{\small (b) system throughput}
\end{minipage}
\caption{The number of completed flows and system throughput under different SI cacellation levels.} \label{fig:under different SI cacellation levels}
\vspace*{-3mm}
\end{figure}
In this case, the contention threshold $\sigma$ is set to 0.001, and the SI cancelation level parameter $\beta$ is uniformly distributed between $2-4$. The simulation results are shown in Figure \ref{fig:under different number of flows}. We can find that the Proposed-FD algorithm always shows superior performance compared with other algorithms, and the more the number of flows, the more obvious the advantages of the Proposed-FD. Compared with the HD algorithms, the Proposed-FD allows simultaneous transmission and reception at the same BS. In fact, the Proposed-HD algorithm also performs better than MQIS. This is because when deciding whether or not to schedule a flow, we consider whether adding the flow can improve the system throughput. This makes each flow be scheduled at a higher rate, and thus the QoS requirements of flows can be achieved more quickly. As for the FDP algorithm, the number of completed flows for it is not large enough. This is mainly because the issue they investigate and the optimization goal are different from ours. It is for the network where the TS resources are sufficient and aims at accomplishing all the transmissions with the minimum time. Therefore, it is not suitable for the investigated problem in this paper that maximizing the completed flows in limited time. Moreover, with the increase of the total number of flows, the number of completed flows for FDP doesn't increase significantly. This is because in FDP, only when all the flows scheduled together in one phase are completed, can we start the next phase and make a new scheduling decision. As a result, when some flows are completed quickly, due to the lack of scheduling of new flows, a large amount of TS resources are wasted. Thus, in limited time, the number of completed flows has almost no change. However, the system throughput of FDP is higher than other HD algorithms. This is because FDP prefers the flows that occupy more TSs. These flows usually have higher QoS requirements (i.e. the minimum throughput requirement), so even the number of completed flows is small, the system throughput is still high. In particular, when the number of flows is 90, the Proposed-FD improves the number of completed flows by 30.1\% compared with Proposed-HD and improves the system throughput by 34.1\% compared with FDP.

\subsubsection{Under different SI cancelation levels}
For the two FD algorithms (the Proposed-FD and FDP), the SI cancelation level $\beta$ has an obvious impact on the performance. Thus, we simulate the performance under different magnitudes of $\beta$, as shown in Figure \ref{fig:under different SI cacellation levels}. The abscissa x is the magnitude of $\beta$. For example, when x = 2, $\beta$ is uniformly distributed in $2\times10^2-4\times10^2$. In this case, the total number of flows is 90, and $\sigma$ = 0.001. We can find that the performance of the Proposed-FD is better when $\beta$ is smaller, that is, when the SI cancelation level is higher. As $\beta$ becomes larger, the performance of the Proposed-FD gradually deteriorates. In particular, when $\beta$ reaches $10^4$ magnitude, the Proposed-FD has the same performance as the Proposed-HD. This tells us that not in any case can the FD communication improve the system performance, and better SI cancelation techniques are needed. The trend of the performance for FDP is similar to the Proposed-FD. However, due to the applicable scenario and optimization goal are different from ours, the performance of FDP is relatively poor.

\begin{figure}[htbp] 
\begin{minipage}[t]{1\linewidth}
\centering
\includegraphics[width=0.9\linewidth]{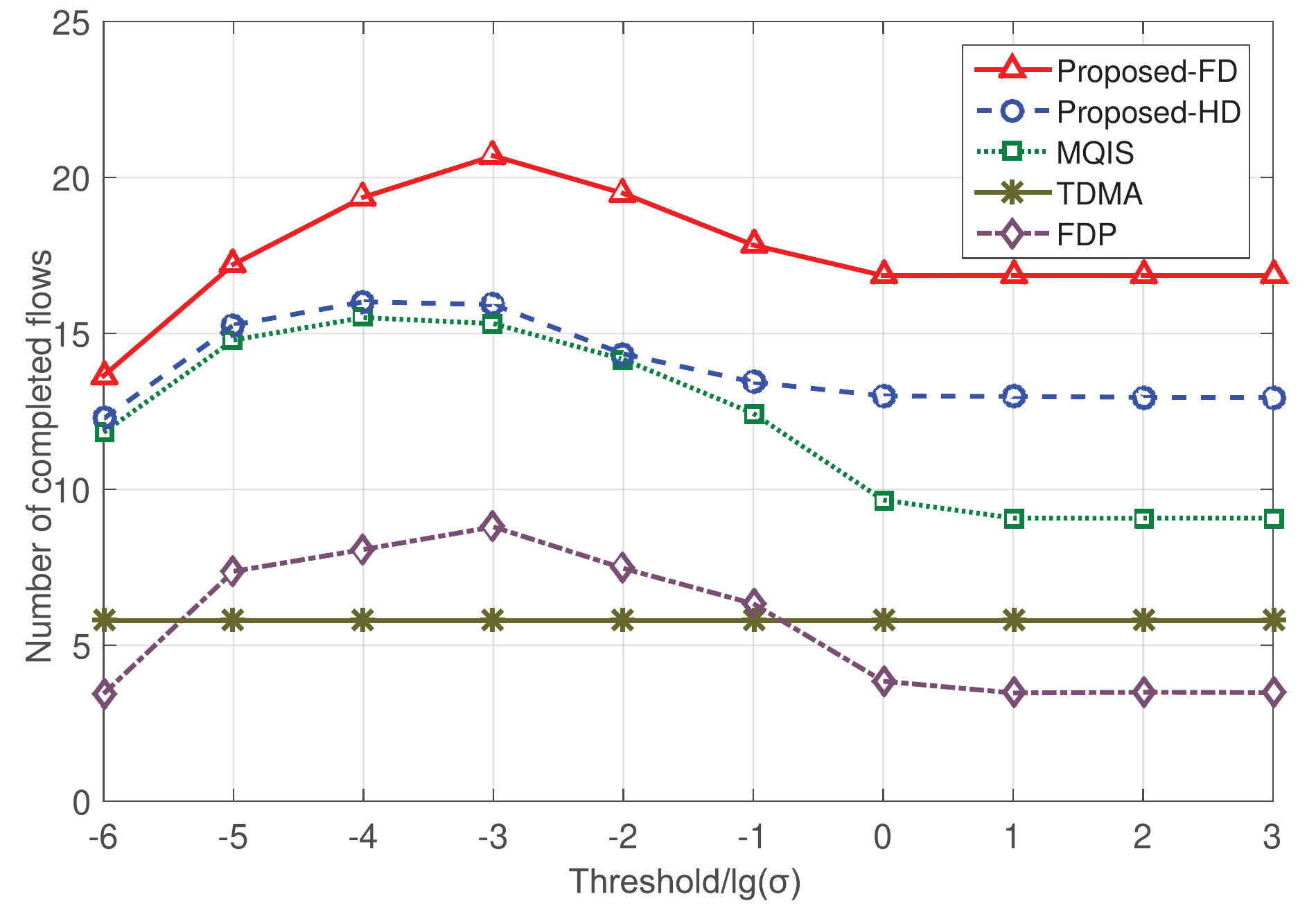}
\centerline{\small (a) number of completed flows}
\end{minipage}\\
\begin{minipage}[t]{1\linewidth}
\centering
\includegraphics[width=0.9\linewidth]{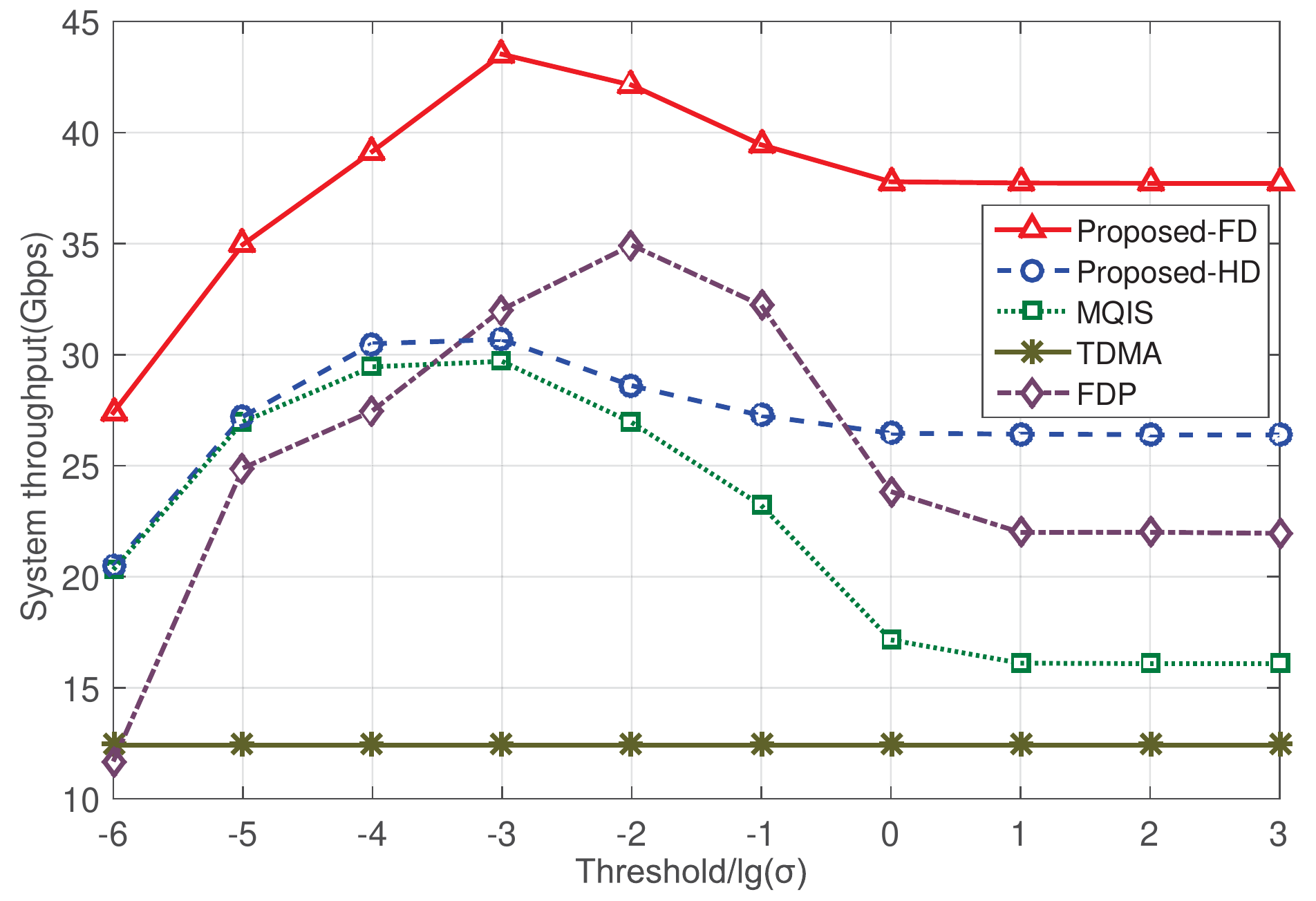}
\centerline{\small (b) system throughput}
\end{minipage}
\caption{The number of completed flows and system throughput under different contention thresholds.} \label{fig:under different contention thresholds}
\vspace*{-3mm}
\end{figure}

\subsubsection{Under different contention thresholds}
To study the impact of contention threshold $\sigma$ on the performance, we simulate the two metrics under different $\sigma$, as shown in Figure \ref{fig:under different contention thresholds}. The abscissa x represents the magnitude of $\sigma$. For example, x = -3 means $\sigma = 10^{(-3)}$. In this case, the number of flows is 90, and $\beta$ is uniformly distributed between $2-4$. We can observe that as $\sigma$ increases, in addition to TDMA, the performance of the other four solutions first increases, then degrades and finally almost keep unchanged. This is because when $\sigma$ is small, it is not conducive to concurrent transmissions. When $\sigma$ is greater than a certain threshold (e.g., $10^{(-3)}$ for Proposed-FD), there is severe interference between concurrent flows, which leads to the rate reduction and is harmful to satisfy the QoS requirements. Therefore, to achieve the best performance, we should choose the appropriate threshold. Under the simulation conditions in this paper, we choose $\sigma=10^{(-3)}$. Specifically, when $\sigma=10^{(-3)}$, the Proposed-FD improves the number of completed flows by 29.9\% compared with Proposed-HD and improves the system throughput by 35.9\% compared with FDP. Although FDP doesn't use the contention graph, we convert its SINR threshold into the contention threshold, so its performance also varies with $\sigma$.

\section{Conclusion}
In this paper, we propose a QoS-aware full-duplex concurrent scheduling algorithm for mmWave wireless backhaul networks. Considering the FD characteristics and the QoS requirements of flows in the system with limited TS resources, the proposed algorithm exploit the contention graph to find the concurrently scheduled flows and maximize the number of completed flows. Extensive simulations show that the proposed FD scheduling algorithm can significantly increase the number of completed flows and the system throughput compared with other scheduling schemes. In addition, the effects of SI cancelation level and contention threshold on the performance are also simulated to guide a better scheduling.

In the future work, we will also consider the blockage problem in mmWave communications into problem, and propose a robust scheme for the mmWave full-duplex backhaul network.


\end{document}